\documentclass[twocolumn,pra,showpacs,amsmath,superscriptaddress]{revtex4-2}
\usepackage{epsfig,graphicx}
\usepackage{bookmark}
\usepackage{braket}
\usepackage{xcolor}

\newcommand{\be}{\begin{equation}}
\newcommand{\e}[1]{\label{#1}\end{equation}}
\def\bea{\begin{eqnarray}}
\def\ea#1{\label{#1}\end{eqnarray}}

\def\ee{\end{equation}}
\def\eea{\end{eqnarray}}
\def\bes#1{\begin{subequations}\label{#1}}
\def\ese{\end{subequations}}

\begin{document}

\title{Reply to Comment(s) to ``Consequences of the single-pair measurement of the Bell parameter''}

\author{Marco Genovese}
\affiliation{INRIM - Istituto Nazionale di Ricerca Metrologica, strada delle Cacce, 91 10135 Torino, Italy.}
\affiliation{INFN, sede di Torino, via P. Giuria 1, 10125 Torino, Italy.}
\author{Fabrizio Piacentini}
\affiliation{INRIM - Istituto Nazionale di Ricerca Metrologica, strada delle Cacce, 91 10135 Torino, Italy.}

\date{\today}

\begin{abstract}
We reply to the comments made by M. Kupczynski and J. P. Lambare on our recent paper titled ``Consequences of the single-pair measurement of the Bell parameter'' [\emph{Phys. Rev. A} \textbf{111}, 022204 (2025)], questioning our claims on the implications on Quantum Mechanics interpretations (and alternative theories) generated by the results of a recent experiment [\emph{Quantum Sci. Technol.} \textbf{9}, 045027 (2024)] demonstrating the possibility of measuring the entire Bell parameter on a single entangled pair.
\end{abstract}

\maketitle




M. Kupczynski and J. P. Lambare comment \cite{kup25,lam25} on our recent paper \cite{gen25}, rising some question on our interpretation of the results presented in Ref. \cite{vir24}. 
In the following we discuss how their questions on our paper \cite{gen25} (and, indirectly, on the meaning of the result obtained in our previous experimental work \cite{vir24}, on which Ref. \cite{gen25} is based) are not correct.\\

In Kupczynski's comment \cite{kup25} to Ref. \cite{gen25} it is claimed that ``the parameter measured in Virz\`{i} \emph{et al.} experiment it is not the Bell parameter $\mathcal{S}$ which was discussed and estimated in many loophole free tests'' and, therefore, this experiment ``neither challenges our understanding of Bell Tests nor allows having doubts about Bohr complementarity and contextuality''.
On the one hand, neither in Ref. \cite{gen25} nor in Ref. \cite{vir24} it is claimed that Bohr's complementarity and contextuality are challenged; what it is claimed there is that certain interpretations of Quantum Mechanics can not straightforwardly explain the observed result.
On the other hand, as discussed in the following, the quantity measured by Virz\`{i} \emph{et al.} \cite{vir24} is exactly the Bell parameter estimated in countless Bell inequality tests, i.e.:
\begin{equation}\label{S}
  \mathcal{S}=\left\langle A_1B_1\right\rangle + \left\langle A_1B_2\right\rangle + \left\langle A_2B_1\right\rangle - \left\langle A_2B_2\right\rangle\;,
\end{equation}
being $\left\langle A_1B_1\right\rangle=\mathrm{Tr}\left(\rho \hat{A}_i\otimes\hat{B}_j\right)$ and $i,j=1,2$, obeying the inequality $|\mathcal{S}|\leq2$ for local hidden-variable theories (LHVTs).\\
In contrast, M. Kupczynski affirms that what is measured in the aforementioned experiment is the parameter
\begin{equation}\label{B}
  \mathcal{B}=\left\langle A_1B_1 + A_1B_2 + A_2B_1 -  A_2B_2\right\rangle\;;
\end{equation}
for dichotomic variables $A_i,B_j=\pm1$, ``for all finite samples the estimates of $|\mathcal{B}|$ are smaller or equal to 2, and the
inequality $|\mathcal{B}|\leq2$ cannot be violated''.\\
Of course, from a quantum mechanical perspective, due to linearity there is no actual difference between $\mathcal{B}$ and $\mathcal{S}$, and the same holds for any LHVT (see, e.g., Eq. (6) in Ref. \cite{kup25}) when the hidden variable distribution $\rho(\lambda)$ is the same for all the measurements.
The claim that ``Neither in an ideal EPRB experiment nor in Bell Tests exist joint probability distributions of $(A_1,A_2,B_1,B_2)$'' is an assumption of specific hidden variable models, and it is exactly this assumption that is challenged by the results of Ref. \cite{vir24}, as discussed in Ref. \cite{gen25}.\\
Nonetheless, in both Eq.s \eqref{S} and \eqref{B} incompatible observables appear, thus in usual projective measurement based experiments one has to measure the different $A_iB_j$ on different photon pairs with different settings of polarizers (i.e., ``different contexts'').
This forces the experimenter to measure the $S$ parameter, preventing the direct extraction of $\mathcal{B}$, making the distinction claimed in Ref. \cite{kup25} arise.
From this point, the author (and others) deduces that the hidden variable distribution $\rho(\lambda)$ may differ for different contexts.\\
Actually, several experiments \cite{lun09,rin14,kum17,cal20,mar21,mit07,the16,pia16seq,ave17,kim18,fol21,vir24,atz24,biz25} have shown that having incompatible observables does not entirely prevent measuring them on the same quantum system and state (e.g., a single entangled photon pair).
Indeed, one has to only satisfy Heisenberg uncertainty principle, eventually accepting larger measurement uncertainties on those observables.
This is exactly the case of the experiment in Ref. \cite{vir24}, where multiple weak measurements \cite{aha88} were used to obtain this result by evaluating all the correlation coefficients appearing in Eq. \eqref{S} on the same photon pair (of course, with the aforementioned large uncertainty for each single measurement, to avoid violating the uncertainty principle).
This is allowed by the fact that weak measurements do not collapse the state, but only induce a small decoherence on it, as demonstrated also in Ref. \cite{vir24}.\\
In this experiment, both parties implement two weak measurements in sequence on one of the photon of the entangled pair, by coupling (weakly) the photon polarization and the transverse momentum induced on the photon via birefringent crystals.
For each weak measurement, this generates a spatial walk-off between the two polarization components in one of the two independent transverse directions $x$ and $y$ (being $z$ the photon propagation axis), keeping the two weak measurements occurring on each photon independent of each other \cite{vir24,atz24}.
After the four weak measurements (two per photon) occur, the two photons of each pair are sent to two spatially-resolving detectors.
This allows extracting, for each two-photon coincidence, the $x$-$y$ coordinates of the firing pixel in both detectors, extracting the coincidence-count coordinate tensor $N(x_A, y_A, x_B, y_B)$, with $A,B$ labelling the outcomes of the two experimenters.
The correlations needed to compute both $\mathcal{S}$ and $\mathcal{B}$ are then simply extracted correlating the firing pixel coordinate with the amplitude of the birefringence shift induced by the weak couplings.\\
In this, there is nothing different with respect to the usual Bell inequality experimental tests, except for the fact that each of the four $x$-$y$ coordinates registered for every two-fold coincidence provides independent information on one of the four polarization measurements.
Such a feature straightforwardly allows extracting at once all the $\left\langle A_iB_j \right\rangle$ on the same entangled pair without substantially altering it (although with the large uncertainty typical of weak measurements), with no need for any model-dependent calculation.
From the operational point of view, this approach makes the evaluation of the $\mathcal{S}$ and $\mathcal{B}$ quantities in Eq.s \eqref{S} and \eqref{B} completely equivalent.
The mathematical equivalence is highlighted in Eq.s (A2)-(A5) of Ref. \cite{vir24}; indeed, Eq. (A4) shows how each of the four correlators forming the Bell-CHSH inequality can be \underline{independently} evaluated from the spatial correlations within the coincidence-count coordinate tensor $N(x_A, y_A, x_B, y_B)$ via Eq.s (A2) and (A3).
The fact that in Eq. (A5) the average is ``global'' is only for the sake of readability, since nothing would change averaging each correlation coefficient and then summing them up, reflecting the complete equivalence of the $\mathcal{S}$ and $\mathcal{B}$ parameters in our framework.\\
Therefore, the meaning of the measurement outcome in Ref. \cite{vir24} is in our opinion perfectly clear, although the single-pair Bell-CHSH parameter estimation might lead, in some cases, to single values apparently largely outside the ``physical range'' $-2\sqrt2\leq\mathcal{S}\leq2\sqrt2$, although perfectly compatible with it within the (large) experimental uncertainties, as it can happen for every noisy measurement both in classical or quantum regime \cite{gen25}.
Furthermore, even if the experiment of Ref. \cite{vir24} was not intended to be loophole free, it has already been discussed \cite{gen25} how it could be, at least in principle, modified for this purpose, e.g. with a scheme of how the two experimenters could implement a random choice of settings.\\
What is claimed, among else, in Ref. \cite{gen25} is that the achieved single photon pair (and single entangled state) measurement of the Bell-CHSH parameter casts doubts on the context interpretation discussed by Kupczynski and others (e.g., that different probability spaces for the hidden variables are needed in different contexts), as well as on certain interpretations of quantum mechanics as modal ones.
We do not think Kupczynski's comment \cite{kup25} solves these objections since, as detailed above, it does not correctly grasp what is done in the experiment of Ref. \cite{vir24}.
Indeed, for discussing the experiment he substantially already assumes the validity of his specific interpretation, since to get rid of the objection in Ref. \cite{gen25} it would require presenting a description of what is measured in Ref. \cite{vir24} in a specific interpretation or model, such as the contextuality by default approach.\\

Such a criticism appears to be at the core of another comment \cite{lam25} by J. P. Lambare, who additionally remarks how the $\mathcal{B}$ quantity in Eq. \eqref{B} [that the author claims to be the one actually measured in the experiment of Ref. \cite{vir24}, in place of the traditional Bell-CHSH parameter $\mathcal{S}$ of Eq. \eqref{S}] could be related to a different inequality based on counterfactuality, i.e. the Bell-Stapp inequality \cite{stapp}.
We already discussed the implications of our experimental result on Stapp's proof of Bell inequalities in Ref. \cite{gen25}, stating how, as the author of Ref. \cite{lam25} claims, our results demonstrate violation of the Bell-Stapp inequality.\\
Nevertheless, we underline that the possibility to perform Bell-Stapp inequality violation tests with the setup of Ref. \cite{vir24} is due to its unprecedented capability of perform measurements on all the (incompatible) bases needed for both this inequality and the Bell-CHSH one on each entangled pair without collapsing its state, getting intrinsically rid of any need for counterfactual reasoning on eventual non-occurring measurements.
This does not mean that what we measure in our experiment is $\mathcal{B}$ instead of $\mathcal{S}$; indeed, what we extract from our measurements is exactly the latter.\\

Finally, we object to Kupczynski's claim that Bell inequality tests do not exclude hidden variable models where the hidden variable density probability (Ref.s [11-13,15-19] in Ref. \cite{kup25}) is context dependent.
Indeed, being the one performed in Ref. \cite{vir24} a single pair measurement, it requires a single hidden variable distribution, therefore only the formalization of a hidden variable model able to reproduce the results of this experiment (and, at the same time, of other Bell inequality tests) would overcome this objection.\\

For the aforementioned reasons, we believe that both these comments \cite{kup25,lam25} do not actually affect neither the validity of our experimental result \cite{vir24} nor its implications for quantum mechanics foundations and interpretations \cite{mg,mg2} reported in Ref. \cite{gen25}.

\section*{ACKNOWLEDGEMENTS}

This work was financially supported by the project Qutenoise (call ``Trapezio'' of Fondazione San Paolo) and AQuTE [Ministry of University and Research (MUR), call
``PRIN 2022'', Grant No. 2022RATBS4].


\begin{references}


%
%
%
%
%
%
%
%
%
%
%
\bibitem{kup25} M. Kupczynski, Comment on ``Consequences of the single-pair measurement of the Bell parameter''. ArXiv:2502.11261 (2025).
%
\bibitem{lam25} J.P. Lambare, Comment on ``Consequences of the single-pair measurement of the Bell parameter''. ArXiv:2504.10341 (2025).
%
\bibitem{gen25} M. Genovese \& F. Piacentini, Consequences of the single-pair measurement of the Bell parameter. \emph{Phys. Rev. A} \textbf{111}, 022204 (2025).
%
\bibitem{vir24} S. Virz\'i et al., Entanglement-preserving measurement of the Bell parameter on a single entangled pair. \emph{Quantum Sci. Technol.} \textbf{9}, 045027 (2024).
%
\bibitem{lun09} J.S. Lundeen, A.M. Steinberg, Experimental Joint Weak Measurement on a Photon Pair as a Probe of Hardy's Paradox. \emph{Phys. Rev. Lett.} \textbf{102}, 020404 (2009).
%
\bibitem{rin14} M. Ringbauer, D.N. Biggerstaff, M.A. Broome, A. Fedrizzi, C. Branciard, A.G. White, Experimental Joint Quantum Measurements with Minimum Uncertainty. \emph{Phys. Rev. Lett.} \textbf{112}, 020401 (2014).
%
\bibitem{kum17} A. Kumari, A.K. Pan, P.K. Panigrahi, Joint weak value for all order coupling using continuous variable and qubit probe. \emph{Eur. Phys. J. D} \textbf{71}, 275 (2017).
%
\bibitem{cal20} O. Calder\'{o}n-Losada, T.T. Moctezuma Quistian, H. Cruz-Ramirez, S.M. Ramirez, A.B. U'Ren, A. Botero, A. Valencia, A weak values approach for testing simultaneous Einstein-Podolsky-Rosen elements of reality for non-commuting observables. \emph{Commun. Phys.} \textbf{3}, 117 (2020).
%
\bibitem{mar21} A.C. Martinez-Becerril, G. Bussieres, D. Curic, L. Giner, R.A. Abrahao, J.S. Lundeen, Theory and experiment for resource-efficient joint weak-measurement, \emph{Quantum} \textbf{5}, 599 (2021).
%
\bibitem{mit07} G. Mitchison, R. Jozsa, S. Popescu, Sequential weak measurement. \emph{Phys. Rev. A} \textbf{76}, 062105 (2007).
%
%
\bibitem{the16} G.S. Thekkadath, L. Giner, Y. Chalich, M.J. Horton, J. Banker, J.S. Lundeen, Direct Measurement of the Density Matrix of a Quantum System. \emph{Phys. Rev. Lett.} \textbf{117}, 120401 (2016).
%
\bibitem{pia16seq} F. Piacentini et al., Measuring Incompatible Observables by Exploiting Sequential Weak Values. \emph{Phys. Rev. Lett.} \textbf{117}, 120402 (2016).
%
\bibitem{ave17} A. Avella, F. Piacentini, M. Borsarelli, M. Barbieri, M. Gramegna, R. Lussana, F. Villa, A. Tosi, I. P. Degiovanni, M. Genovese, Anomalous weak values and the violation of a multiple-measurement Leggett-Garg inequality. \emph{Phys. Rev. A} \textbf{96}, 052123 (2017).
%
%
\bibitem{kim18} Y. Kim, YS. Kim, SY. Lee, S. Moon, YH. Kim, YW. Cho, Direct quantum process tomography via measuring sequential weak values of incompatible observables. \emph{Nat. Commun.} \textbf{9}, 192 (2018).
%
\bibitem{fol21} G. Foletto, M. Padovan, M. Avesani, H. Tebyanian, P. Villoresi, G. Vallone, Experimental Test of Sequential Weak Measurements for Certified Quantum Randomness Extraction. \emph{Phys. Rev. A} \textbf{103}, 062206 (2021).
%
\bibitem{atz24} F. Atzori et al., Experimental Test of Nonlocality Limits from Relativistic Independence. \emph{PRX Quantum} \textbf{5}, 040351 (2024).
%
\bibitem{biz25} G. Bizzarri, S. Gherardini, M. Manrique, F. Bruni, I. Gianani, M. Barbieri, Quasiprobability distributions with weak measurements. \emph{Quantum Sci. Technol.} \textbf{10}, 045008 (2025).
%
\bibitem{aha88} Y. Aharonov, D. Z. Albert, L. Vaidman, How the result of a measurement of a component of the spin of a spin-1/2 particle can turn out to be 100. \emph{Phys. Rev. Lett.} \textbf{60}, 1351 (1988).
%
\bibitem{stapp} H. P. Stapp, S-matrix interpretation of quantum theory. \emph{Phys. Rev. D} \textbf{3}, 1303 (1971).

\bibitem{mg} M. Genovese, Interpretations of Quantum Mechanics and Measurement Problem. \emph{Adv. Sci. Lett.} \textbf{3}, 249 (2010).

\bibitem{mg2} W. Myrvold, M. Genovese, A. Shimony, Bell's Theorem. In ``The Stanford Encyclopedia of Philosophy'' (Eds.: Edward N. Zalta and Uri Nodelman, Metaphysics Research Lab, Stanford University, 2024)

%
%
%
%
%
%
%
%
%
%
%
\end{references}
\end{document}